\documentclass[twocolumn,prb,preprintnumbers,amsmath,amssymb]{revtex4}

\usepackage{graphicx}
\usepackage{dcolumn}
\usepackage{bm}
\usepackage{hyperref}

\renewcommand{\vec}[1]{\mathbf #1}
\newcommand{\mv}[1]{\langle#1\rangle}                           
\newcommand{\Mv}[1]{\left\langle#1\right\rangle}                
\newcommand{\Abs}[1]{\left| #1\right|}                          
\newcommand{\abs}[1]{| #1|}                                     

\begin{document}

\title{Generating multi-chain configurations of an inhomogeneous melt from the knowledge of single-chain properties}

\author{Martin H{\"o}mberg}
\author{Marcus M{\"u}ller} \email{mmueller@theorie.physik.uni-goettingen.de}
\affiliation{Institut f{\"u}r Theoretische Physik, Georg-August Universit{\"a}t, 37077 G{\"o}ttingen, Germany}
\begin{abstract}
Mean-field techniques provide a rather accurate description of single-chain conformations in spatially inhomogeneous polymer systems containing interfaces or surfaces. Intermolecular correlations, however, are not described by the mean-field approach and information about the distribution of distance between different molecules is lost. Based on the knowledge of the exact equilibrium single-chain properties in contact with solid substrates, we generate multi-chain configurations that serve as nearly equilibrated starting configurations for molecular dynamics simulations by utilizing the packing algorithm of Auhl and co-workers [J. Chem. Phys. {\bf 119}, 12718 (2003)] for spatially inhomogeneous systems, i.e., a thin polymer film confined between two solid substrates. The single-chain conformations are packed into the thin film conserving the single-chain properties and simultaneously minimizing local fluctuations of the density. The extent to which enforcing the near-incompressibility of a dense polymer liquid during the packing process is able to re-establish intermolecular correlations is investigated by monitoring intermolecular correlation functions and the structure function of density fluctuations as a function of the distance from the confining solid substrates.  
\end{abstract}

\maketitle
\section{Introduction}
\label{introduction}
Many modern polymer materials consist of multiple components.\cite{Bates99,Ruzette05} While the material appears homogeneous on a macroscopic length scale, the different constituents often do not mix microscopically; domains of micrometer size form and they are separated by internal interfaces. The properties of these materials arise from a complex interplay between the characteristics of the individual components in the bulk and interfacial properties.\cite{Mat1,Mat2}

Therefore, abiding interest has been directed towards investigating inhomogeneous polymer systems, e.g., surfaces of polymeric materials and polymer-solid contacts, interfaces in homopolymer blends, or microphase-separated morphologies in block copolymers. Computer simulations are an ideal tool to provide simultaneous insights into (i) the molecular conformations of the long flexible macromolecules at interfaces or surfaces,  (ii) thermodynamic or mechanical properties (e.g., interfacial tension) and (iii) the single-chain dynamics in the inhomogeneous environment and the kinetics of structure formation.

The simulation of large systems of long chain molecules, however, poses computational challenges and alternative approaches, like the self-consistent field theory or polymer density functional theory, have been successfully applied to study inhomogeneous polymer systems.\cite{Helfand72b, Scheutjens79, Hong81, Shull90, Matsen94, Schmid95, Werner99d, Tyler03, Tyler05, Matsen06} These mean-field techniques approximate the problem of interacting chains by the much simpler one of a single molecule in an external field that, in turn, mimics the interactions with its surrounding. Thus, significantly larger systems of longer chains can be described by mean-field theory than by computer simulations. In dense systems of long flexible molecules the mean-field approach yields a rather accurate description of long wave length properties.\cite{deGennesBook} The accuracy has been demonstrated by mapping the representations used in mean-field theories and computer simulations and quantitatively comparing the results of both techniques.\cite{Muller95b, Werner97,  Muller98b, Muller00f, Muller03d} In particular, the mean-field theory provides an excellent, quantitative description of the molecular conformations in an inhomogeneous system.

Given an accurate mapping between the  representations used in mean-field theories and computer simulations, it is tempting to utilize the information gained by mean-field theory to generate large, equilibrated multi-chain conformations that can be studied by computer simulations (e.g., to investigate the local dynamics of the molecules at an interface).\cite{Sewell07} However, there is a rather fundamental problem: The {\em mean-field theory ignores all intermolecular correlations}, i.e. approximating the multi-chain system by that of a single chain in an external field, all information about the relative distances between molecules is lost.  

In order to illustrate this problem let us consider the simpler case of a homogeneous dense polymer melt where the predictions of the mean-field theory are trivial. Within mean-field theory the chain conformations are unperturbed. If one generated single-chain conformations according to the pre-defined bulk distribution and placed them randomly into a simulation cell, the resulting multi-chain configuration would not represent a dense melt at all. The collective structure function of density correlations would be identical to the single-chain structure factor and the compressibility would be that of an ideal gas of chains rather than adopting the low value characteristic for a dense polymer liquid.\cite{deGennesBook} Likewise, the correlations between segments belonging to different molecules would be trivial instead of exhibiting a correlation hole that is the hallmark of intermolecular pair correlations in a melt.\cite{deGennesBook} Since the spatial extent of the correlation hole is long-ranged, the chains have to diffuse a distance of the order of their end-to-end distance, $R_e$, in order to build-up the intermolecular correlations. The necessary time, $\tau_R$, increases with chain length and much of the benefit of using the information of the mean-field theory would be lost. Moreover, the correlation hole of the intermolecular correlation function also gives rise to important, long-ranged intramolecular correlations in semi-dilute solutions \cite{Muller00j} and melts\cite{Semenov03} that can be observed in the power-law decay of the bond-bond correlation function along the chain contour\cite{Wittmer04,Wittmer07,Wittmer07c} or a hump in the Kratky-plateau of the single-chain structure factor.\cite{Muller00j,Beckrich07} It is important to emphasize that both, intra- and intermolecular correlations on the scale $R_e$ do not explicitly depend on the strength of the excluded volume interactions or the details of the interaction potentials but are determined by the polymer density and chain extension.\cite{Wittmer04,Wittmer07,Wittmer07}

Problems due to randomly placing Gaussian chains in a simulation box have already been observed.\cite{Brown94, Auhl03} It was found that after randomly placing chains in a simulation cell and equilibrating the resulting multi-chain configuration by molecular dynamics simulations, the chain conformations initially swell and a time of the order of the Rouse time, $\tau_R$, is required to achieve equilibrium. Furthermore, Auhl {\em et al.}~demonstrated that if the chains are placed in the simulation box such that the density is uniform, much of the swelling of the conformations will be avoided and equilibrium will be achieved faster.\cite{Auhl03}

In the present work, we test the hypothesis that the incompressibility constraint, i.e. the strong reduction of long wave length density fluctuations in a dense polymeric fluid, is sufficient to generate the intermolecular correlations that are ignored by mean-field theory. To this end, we adopt the packing algorithm of Auhl and co-workers \cite{Auhl03} for homogeneous systems and apply it to a thin polymer film confined between two structureless solid substrates. This simple spatially inhomogeneous system serves as a testing bed because it has been studied by a variety of self-consistent field and polymer density functional theory approaches.\cite{Yeth95, Yethiraj98, Muller00f, Yu02b, Muller03d, bryk04, Wu06, Wu07} The method can be readily generalized to other types of inhomogeneities. Moreover, rather than using Gaussian chain conformations, we employ single-chain conformations that are extracted from an equilibrated melt.\cite{Muller97, Muller00f, Muller03d} These conformations are not strictly Gaussian on large length scales but they already include the non-trivial long-ranged intramolecular correlations.

The manuscript is arranged as follows: In the next section \ref{model}, we describe the coarse-grained polymer model, the simulation techniques and the packing algorithm. In Sec.~\ref{results}, we investigate how close the so-generated multi-chain configurations of a thin polymer film are to equilibrium. Single-chain conformations, intra- and intermolecular correlations as well as density fluctuations will be considered. The paper concludes with a brief summary and outlook.
      
\section{Model and Technique}
\label{model}

\subsection{Model}
We consider a widely utilized coarse-grained bead-spring model for polymers,\cite{grest, LJ_pot2, Baschnagel05} which has successfully been applied  to many different physical situations and for which many data are readily available for reference. The generic coarse-grained model only includes the relevant interactions that are necessary to describe the universal behavior of long, flexible macromolecules in a dense solution. Harsh repulsive excluded volume interactions and longer-ranged, attractive van-der-Waals attractions act between all segments of all chains and are modeled by a spherically truncated and shifted Lennard-Jones (LJ) potential,
\begin{equation}
\label{lj-potential}
U_{\text{LJ}}(r) = 
\begin{cases}
4\varepsilon \left[ \left(\frac{\sigma}{r}\right)^{12} - \left(\frac{\sigma}{r}\right)^{6} + \frac{127}{16384}\right], & r \le r_c \\
0, & r > r_c
\end{cases}
\end{equation}
$\varepsilon$ and $\sigma$ define the units of energy and length, respectively. The cutoff radius, $r_c=2\times 2^{1/6} \sigma$, is given by twice the position of the minimum in Eq.~\eqref{lj-potential}. Additionally, the connectivity along the macromolecule is modeled by a finitely extensible nonlinear elastic (FENE) potential that acts between bonded neighbor segments along the same polymer chain,
\begin{equation}
\label{fene-potential}
U_{\text{FENE}}(r) = 
\begin{cases}
-\frac{kR_0^2}{2}\cdot \ln{\left[1-\left(\frac{r}{R_0}\right)^2\right]}, & r \le R_0 \\
\infty, & r > R_0
\end{cases}
\end{equation}
where $R_0 = 1.5\,\sigma$ is the maximum extension of a bond and $k=30\,\varepsilon/\sigma^2$ is the spring constant.

Polymer segments interact with the unstructured solid substrate via an integrated Lennard-Jones 3-9-potential
\begin{equation}
\label{wall-potential}
U_{\text{wall}}(z) = \varepsilon \varepsilon_W \left[ \left(\frac{\sigma}{z}\right)^{9} - \left(\frac{\sigma}{z}\right)^{3}\right],
\end{equation}
where the dimensionless Hamaker constant $\varepsilon_W=3.0$ is a measure for the attraction between solid substrate and polymer melt.

Immediately after rearranging the chains by means of the packing algorithm, which will be described below, there still remains some overlap between different segments. In this case, the use of Eq.~\eqref{lj-potential} for non-bonded interactions leads to numerical instabilities. At this initial stage, we therefore employ for all non-bonded interactions a potential with a maximum force at short distances $r<r_{\text{fc}}$. 

\begin{equation}
\label{ljfc-potential}
U_{\text{FC}}(r) = \begin{cases}
\left(r-r_{\text{fc}}\right)\cdot U'_{\text{LJ}}(r_{\text{fc}}) + U_{\text{LJ}}(r_{\text{fc}}), & r < r_{\text{fc}} \\
U_{\text{LJ}}(r), & r \ge r_{\text{fc}}
\end{cases}
\end{equation}
In the limit  of vanishing $r_{\text{fc}} \to 0$, the potential converges to the full interaction potential.\cite{Auhl03}

In the following, we study a monodisperse melt of $n$ polymers that are comprised of $N$ effective segments per chain in the canonical ensemble. We use a temperature of $k_BT/\varepsilon = 1.68$ where $k_B$ denotes Boltzmann's constant. The melt is confined between two impenetrable unstructured solid substrates parallel to the $xy$-plane. The solid substrates are located at $z=-L_z/2$ and $z=L_z/2$. Periodic boundary conditions are applied in the two other directions, $x$ and $y$. The mean segment number density, $\rho$, as well as the other parameters are compiled in table \ref{tab1}. 

\begin{table}[tb]
\begin{tabular}{|l|l|l|l|l|l|l|l|l|}
\hline
& $N$ & $n$ & $L_z/\sigma$ & $\rho\sigma^3$ & $D\tau/\sigma^2$ & $R_e^2/\sigma^2$ & $R_g/\sigma$ & $\tau_R/\tau$ \\
\hline
I   &  64  &  300  & 31.56  & 0.611 & $4.36\cdot 10^{-3}$ & 104 & 4.19 & 805 \\
II  & 128  &  230  & 34.62  & 0.710 & $1.27\cdot 10^{-3}$ & 209 & 5.88 & 5558 \\
III & 256  &  230  & 43.61  & 0.710 & $2(1)\cdot 10^{-4}$ & 446 & 8.74 & 8(4)$\cdot10^4$ \\
\hline
\end{tabular}
\caption{\label{tab1}Parameters of the simulations. $N$: number of effective segments (interaction centers) per chain. $R_e$: mean-squared end-to-end distance. $D$: self-diffusion coefficient of the molecules' center of mass. $\tau_R$: Rouse times calculated via $\tau_R = R_e^2/(3\pi^2 D)$.\cite{DoiEdwards}}
\end{table}

Two different methods have been utilized to equilibrate the configurations: On the one hand, a Monte Carlo (MC) program with a mixture of both, slithering snake and random local moves, is employed because these MC moves can quickly equilibrate even long chains at moderate density. On the other hand, we use a parallel Molecular Dynamics (MD) program with Nos{\'e}-Hoover thermostat to simulate at constant temperature.\cite{NOSE84, HOOVER85} Within this technique, the system is coupled to a heat bath, which is represented by one additional degree of freedom, $\eta$, with canonically conjugated momentum, $p_\eta$. The concomitant equations of motion are
\begin{subequations}
\label{eom}
\begin{eqnarray}
\mathbf{\dot r}_i = \frac{\mathbf{p}_i}{m}, && \quad
\mathbf{\dot p}_i = \mathbf{F}_i - \frac{p_\eta}{Q} \mathbf{p}_i, \\
\dot\eta = \frac{p_\eta}{Q}, && \quad
\dot p_\eta = \sum\limits_{i=1}^{nN} \frac{\mathbf{p}_i^2}{m}-g k_B T.
\end{eqnarray}
\end{subequations}
$\eta$ fluctuates around zero; its ``mass", $Q$, is chosen as to realize a canonical ensemble. $m$ is the mass of a segment and we set $m=1$ in the remainder of this article. $g=3nN$ denotes the number of degrees of freedom.  The equations of motion \eqref{eom} were integrated using the velocity Verlet algorithm \cite{frenkel-smit,tildesley} with a time step of $\Delta t=0.002\,\tau$, where $\tau=\sigma\sqrt{m/\varepsilon}$ denotes the time unit in Lennard-Jones parameters.

\subsection{Packing process}
\label{packing-process}
In this section, we describe our packing algorithm, which we have used to create polymer melts of long chains that are already very close to equilibrium. This algorithm is based on an idea of Auhl and co-workers.\cite{Auhl03} Here, we generalize it to spatially inhomogeneous systems. The underlying hypothesis is to create intermolecular correlations by enforcing incompressibility of the dense polymer solution. The strategy of the packing algorithm consists in reducing local fluctuations of the segment density without corrupting the intramolecular correlations. This algorithm has been implemented as a Monte Carlo program performing trial moves on {\em a priori} equilibrated chain conformations, which are treated as rigid bodies.

The equilibrated polymer melt in contact with the solid substrate exhibits $z$-dependent intra- and intermolecular correlations on different length scales.\cite{Muller02c,Cavallo05} The spatial inhomogeneity makes correlation effects also visible in the spatial dependence of single-chain properties (e.g., the segment density distribution). On the one hand, there are intricate packing effects of the fluid of segments that manifest themselves in oscillations of the density profile as a function of the distance, $z$, from the solid substrate. In a dense melt, this local arrangement of segments decays on the length scale of the segment size or the screening length, $\xi$, of density fluctuations, which is microscopic. Our packing algorithm does not aim at generating this short-ranged fluid ordering because this local structure on the segmental scale equilibrates on time scales that are fast compared to the Rouse time of the macromolecules. On the other hand, the presence of the solid substrate influences the polymer structure over distances up to the molecular size. In this region -- denoted interphase -- the polymer chains align with the solid substrate, the back-folding of the conformations deepens the correlation hole of the intermolecular pair correlation function and the density of the molecules' center-of-masses depends non-trivially on the distance from the solid substrate. These effects have to be captured by the packing algorithm because their equilibration requires the molecules to diffuse a distance of the order of the molecular extension and the concomitant time scale is set by the Rouse-time.\cite{DoiEdwards}

Our packing algorithm requires (i) equilibrated single-chain conformations of a bulk system that do incorporate the non-trivial long-ranged intramolecular correlations imparted by the incompressibility constraint and (ii) the distribution of the single-chain conformations in the spatially inhomogeneous system. The latter property can be obtained with high accuracy by self-consistent field or polymer density functional calculations evaluating the single-chain partition function via a partial enumeration over a large ensemble of bulk single-chain conformations. 

In order to assess the quality of the packing algorithm, however, we adopt in the following a different procedure that does not invoke the additional approximations of the self-consistent field or polymer density functional calculations. We equilibrate a dense polymer melt in the presence of the solid substrate. This equilibrated multi-chain configuration serves (i) as a reference to which we compare the result of the packing process and (ii) it provides {\em exact} information about the distribution of single-chain conformations in the spatially inhomogeneous system. We extract single-chain conformations at a given distance from the spatial inhomogeneity from equilibrated melt configurations (at different times) and our packing algorithm re-assembles them to a multi-chain configuration. Therefore, the deviations of the multi-chain configuration from the equilibrated melt solely stem from the packing algorithm.

In the first stage, we generated an initial multi-chain configuration by randomly drawing equilibrated molecular conformations from a large ensemble of single-chain conformations. The distance from the spatial inhomogeneity (i.e., the $z$-coordinate) and the orientation of each single-chain conformation in the newly assembled configuration is identical to the value in the equilibrated configuration and the density profiles of both configurations are also statistically identical. We emphasize that only information about single-chain properties,  which can be obtained from a mean-field theory, enter the packing process. This first stage generates an initial configuration that reproduces the single-chain properties of the equilibrated melt (e.g., the molecules' center-of-mass distribution) but the intermolecular correlations parallel to the solid substrate are that of an ideal gas of polymers.  

In the following packing procedure, single-chain conformations are treated as rigid bodies (i.e., the intramolecular correlations are frozen) and the distance to the (nearest) solid substrate as well as the orientation of the molecule is kept fixed. In order to introduce intermolecular correlations in the direction parallel to the spatial inhomogeneity, we minimize local density fluctuations. To this end, we define a local number of neighbors, $n_{i,s}$ of the $s^{\rm th}$ segment of polymer $i$ by counting the number of segments within a sphere of radius $r_p$ around the reference segment.\cite{Auhl03} This procedure guarantees translational invariance and isotropy in the bulk. In a spatially inhomogeneous system, however, the local number of neighbors and its fluctuations depend on the distance, $z$, from the solid substrate. In practice, we discretize the system into $2n_b+1$ bins such that there are $n_b$ bins of constant width $\Delta_k$ on each side. For the multi-chain configuration, we define the local number of neighbors in layer $z_k$ with $k=1,\cdots,2n_b+1$ according to
\begin{equation}
n(z_k) = \frac{\int {\rm d}z\;\sum_{i=1}^n\sum_{s=1}^N n_{i,s} \delta(z-z_i)}{\int {\rm d}z\;\sum_{i=1}^n\sum_{s=1}^N \delta(z-z_i)}
\end{equation} 
where the integral runs over the $k^{\rm th}$ bin. The variance of the local number of neighbors is a measure of density fluctuations and we minimize it for each layer parallel to the solid substrate. Thus, the target function for the second step of the packing process is given by
\begin{equation}
\label{Ez}
E=\sum_{k=1}^{2n_b+1} \Delta_k E(z_k)\qquad \mbox{with}\quad E(z_k)=\mv{n^2}(z) - \mv{n}^2(z).
\end{equation}

To preserve the structure of the inhomogeneous system, the distance between every segment and the inhomogeneity must not change during the second stage of the packing process. The inhomogeneity under consideration is both invariant under translations in the parallel directions, $x$ and $y$, and rotations around the perpendicular $z$-axis. Additionally, it is invariant under reflections at the middle plane of the thin film. In order to reduce local density fluctuations and simultaneously preserve the structure in $z$-direction, we utilize the following set of moves in the second stage: (i){\em Translation} of a single chain parallel to the solid substrates. (ii) {\em Rotation} of a single chain by an arbitrary angle along an axis perpendicular to the solid substrates. (iii) {\em Reflection} of a chain by the middle plane of the system. (iv) {\em Replacement} of a chain in the system by a different chain at a different position according to the mean-field prediction of the single-chain properties. Such a change of a single-chain conformation is accepted if it will decrease the target function, $E$.

The typical values we have employed are $n_b=5$ resulting in $\Delta=3.0\,\sigma$. The second stage of the packing process started with a value of $r_p=4.0\,\sigma$, which was successively decreased in steps of $\Delta r_p=1.0\,\sigma$ down to $r_p=2.0\,\sigma$. The typical runtime of the packing process for the system with $n=300$ and $N=64$ with a total number of $4\cdot 10^6$ MC steps on a single Intel Xeon 5160 CPU is roughly $3\,$h.

After this second stage of the packing process, the multi-chain configuration is utilized as starting configuration of a MD run. Since, immediately after the second stage, there still remain some local overlaps of the segments, we cannot use the full interaction potential but rather employ the linearized potential \eqref{ljfc-potential} for non-bonded interactions. The initial value of $r_{\text{fc}}=0.9\,\sigma$ is gradually decreased by $0.001\,\sigma$ every $20\,\tau$ until a final value of $r_{\text{fc}}=0.82\,\tau$ was reached. This ``push-off'' is very effective in removing local segmental overlaps. This third stage completes the generation of an overlap-free, inhomogeneous multi-chain configuration for our coarse-grained bead-spring model.

In order to assess how close to equilibrium the so-generated configuration is, we studied the configurations with the full interactions between all segments for additional $3400\,\tau$ by MD simulations. Initially, we assigned random velocities to each segment according to the Maxwell-Boltzmann statistics such that the temperature in the initial state already matched the target temperature.

\section{Results and Discussion}
\label{results}

\subsection{Packing Process}
We apply the packing process described in Sec.~\ref{packing-process} to confined polymer melts, which are comprised of molecules with different chain lengths, $N=64, 128, 256$. In this subsection, we describe the impact of this packing scheme on the properties of the polymer melt with $N=64$ by inspecting the correlation hole and the collective structure function.

The collective structure function of the density is given by
\begin{equation}
\label{Gq}
G(q_\parallel) = \frac{1}{nN}\Mv{\Abs{\sum\limits_{i=1}^n\sum\limits_{s=1}^N e^{i\vec q_\parallel \vec r_{i,s}}}^2}.
\end{equation}
It is a measure of the strength of density fluctuations on all length scales. In Eq.~\eqref{Gq}, $\vec q_\parallel$ denotes a wave vector parallel to the solid substrate and $q_\parallel \equiv \abs{\vec q_\parallel}$. Due to the periodic boundary conditions in the two parallel directions, the components of $\mathbf{q}_\parallel$ are restricted to $q_{\parallel,i} = 2\pi n_i/L_i$ with $n_i=0,1,2,\dots$. Wave vectors are measured in units of the molecules' radius of gyration, $R_g$.

\begin{figure}[tb]
\includegraphics[clip,width=1.\linewidth]{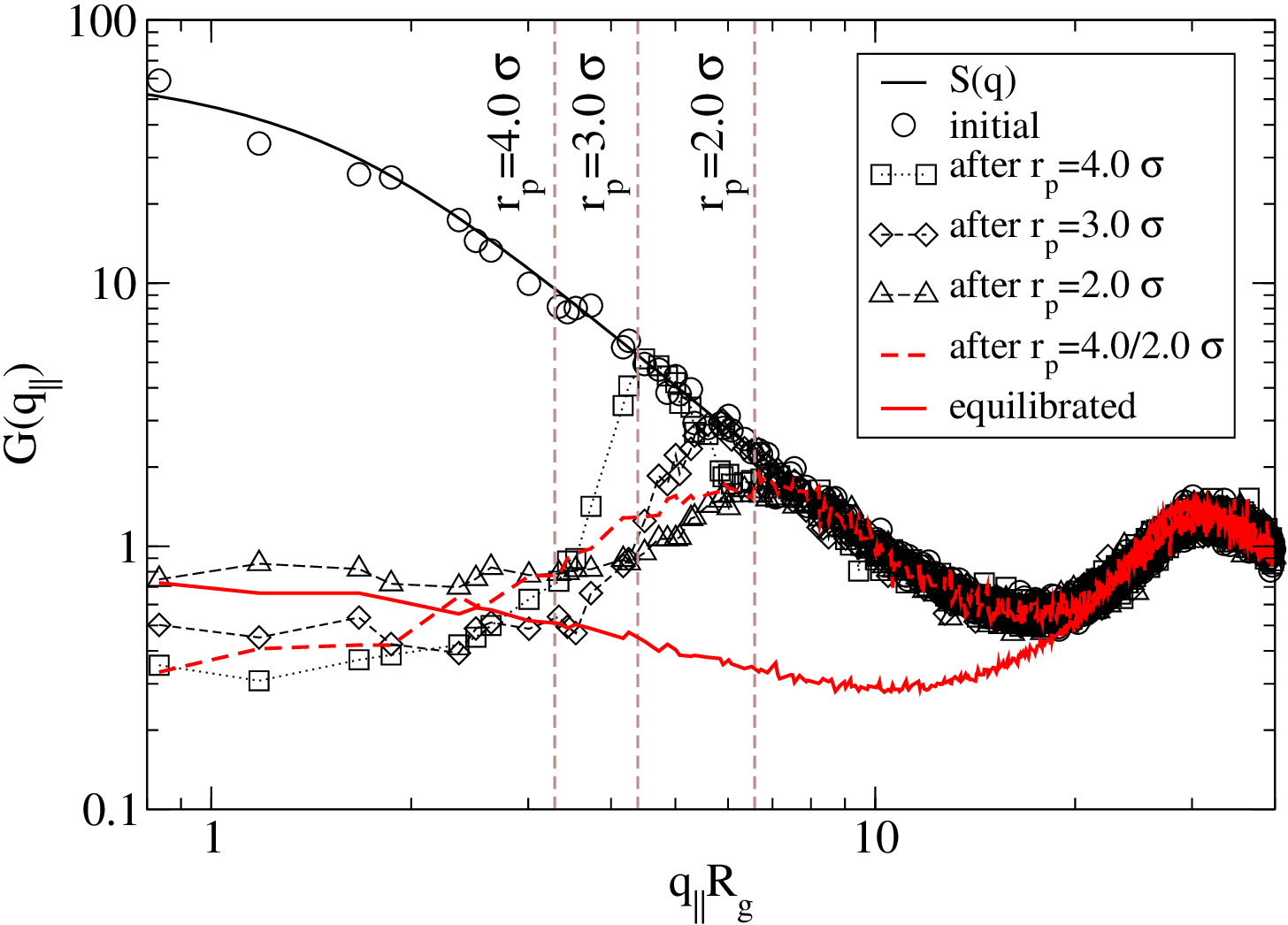}
\caption{\label{fig:gq64_all}Comparison of the collective structure function, $G(q_\parallel)$ ($N=64$, $n=300$, $\rho\sigma^3=0.611$), of the system in equilibrium with the packed system before, after the completion of each packing step with a certain radius $r_p$, and after application of mixed moves. $S(q)$ denotes the single-chain structure factor.}
\end{figure}

In Fig.~\ref{fig:gq64_all}, we present $G(q_\parallel)$ during the second stage of the packing process with $r_p=4.0\,\sigma, 3.0\,\sigma,$ and $2.0\,\sigma$, respectively. The data after the first stage of the packing process and in equilibrium are also displayed for comparison. The values of $q_\parallel$ corresponding to $2 r_p$ (i.e., the diameters of the sphere around a segment) are indicated by vertical dashed lines in the figure. 

Since the lateral arrangement of single-chain conformations  is random after the first stage of the packing process, $G(q_\parallel)$ is exactly given by the structure function of a single chain
\begin{equation}
\label{Sq}
S(q) = \frac{1}{N}\Mv{\Abs{\sum\limits_{s=1}^N e^{i\vec q\vec r_s}}^2}.
\end{equation}
The peak at large wave vectors, $q_\parallel R_g\approx 32$, is a consequence of the intramolecular packing and reflects the length scale of a single bond. In an equilibrated dense melt, this peak also exists but it stems from both, intra- and intermolecular packing. In our model, there is a small mismatch between the preferred bonded and non-bonded distances,\cite{LJ_pot2} which explains the small difference between this local packing peak after the first stage of the packing and in an equilibrated melt. 

During the second stage of the packing process, density fluctuations on the length scale $r_p$ are reduced and $r_p$ is successively decreased. The impact of the packing algorithm can be observed quite clearly by the presence and shift of the additional peak in $G(q_\parallel)$ at intermediate wave vectors. The packing reduces density fluctuations for $q_\parallel \ll 2\pi/r_p$ but leaves fluctuations on shorter length scales as well as the single-chain structure unaltered. Thus, the collective structure function develops a peak around $q_\parallel^* \approx 2\pi/2r_p$, which shifts to larger wave vectors as we decrease $r_p$. 

After packing with the largest value, $r_p=4.0\,\sigma$, the density fluctuations have been significantly reduced on length scales $q_\parallel R_g < 4$; in fact, they are even smaller than in equilibrium. In the course of the subsequent packing, the fluctuations for small $q_\parallel$ slightly increase again because both long-ranged and shorter-ranged fluctuations are simultaneously  minimized. After packing with $r_p=3.0\,\sigma$ the peak in $G$ has moved to $q_\parallel R_g \approx 5.8$ and its intensity has further decreased. After the final packing with $r_p=2.0\,\sigma$, this peak decreases still further, however, the small length scales remain unaffected by the packing process, i.e. $G(q_\parallel)$ still resembles the single-chain structure factor, $S(q)$, for $q_\parallel R_g > 8$. Any further decrease of $r_p$ does not improve the situation. Instead of moving $G(q_\parallel)$ closer to its equilibrium value, we observe that long wave length density fluctuations increase, i.e., the small $q$-regime is getting worse in favor of the large $q$-regime. Since the relaxation times for large $q$ are small, we decide to terminate the second stage of the packing process at $r_p=2.0\,\sigma$.

A comparison of these structure functions with $G(q_\parallel)$ of an equilibrated melt (cf.~Fig.~\ref{fig:gq64_all}) demonstrates that the second stage of the packing process is able to decrease the long wave length fluctuations significantly and that the collective structure function for small values of $q_\parallel$ approaches the equilibrium value.

In passing, we note that the limiting value $G(q_\parallel \to 0)$ is not fixed by the packing process, but can be adjusted to values between that of the second stage of the pacing process at $r_p=2.0\,\sigma$ and that of $r_p=4.0\,\sigma$ by mixing moves with different values of $r_p$. Choosing a ratio of $9:1$ of moves with $r_p=2.0\,\sigma$ and $4.0\,\sigma$, we observe in Fig.~\ref{fig:gq64_all} that the position of the packing peak levels off, however the peak broadens slightly towards smaller values of $q_\parallel$.

\begin{figure}[tb]
\includegraphics[clip,width=1.\linewidth]{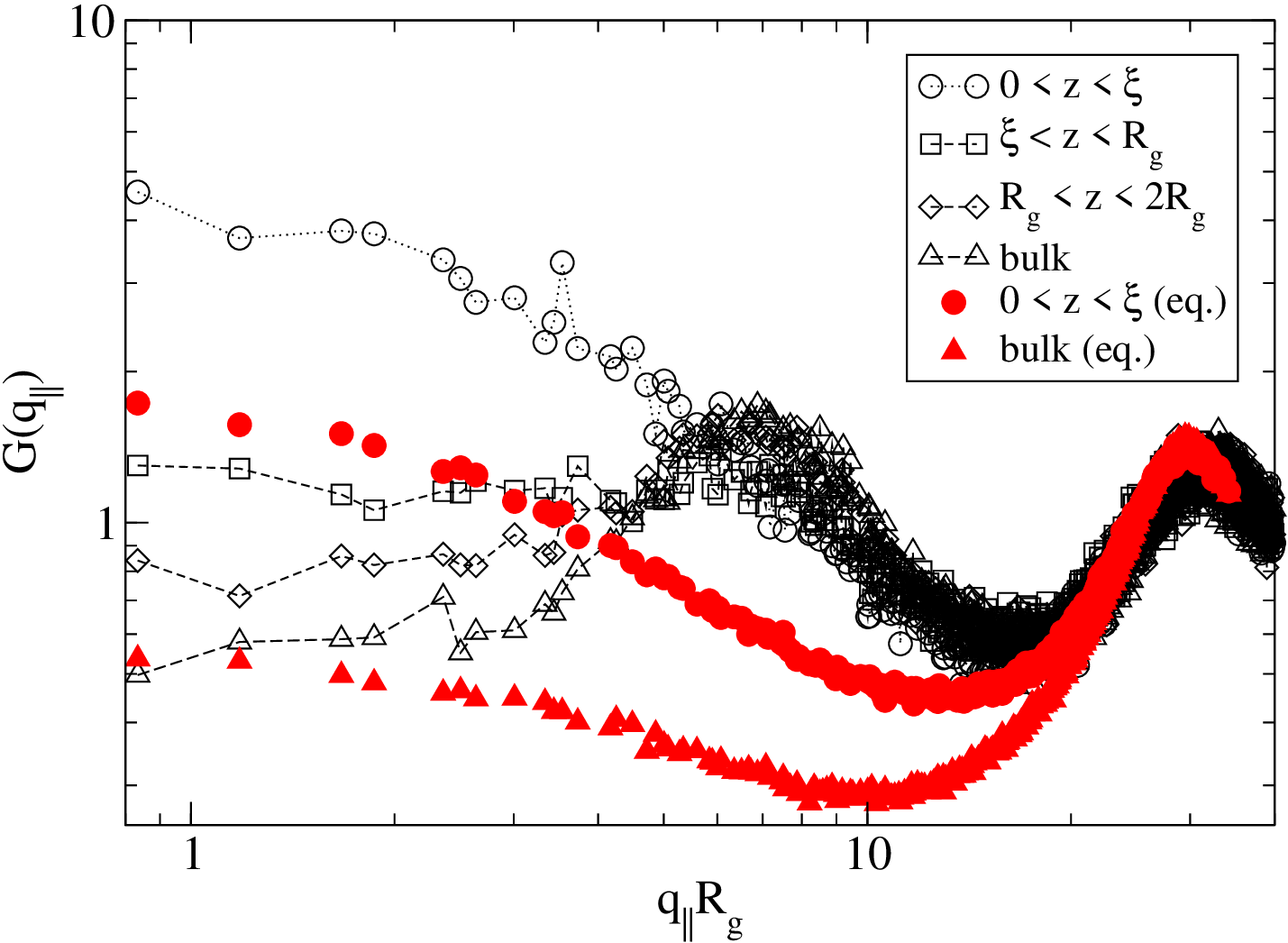}
\caption{\label{fig:gqz64_4M}$G(q_\parallel)$ after packing with $r_p=2.0\,\sigma$ for the intervals specified in the text ($N=64$). The noticeable discrepancy between the equilibrium distribution and that from the packing algorithm at $0<z<\xi$, poses no severe problems, since the distribution of the molecules center-of-mass is correct by construction and the segments will move quickly into this volume.}
\end{figure}

We also investigate the packing as a function of the distance from the spatial inhomogeneity. Since the relevant length scale for local packing is given by the correlation length $\xi$ ($\xi\approx2.5\,\sigma$ for $N=64$) and the length scale for intermolecular packing is given by the radius of gyration, $R_g$, we divided the system into four slabs: The first interval from $0<z<\xi$ is close to the solid substrates, the second spans $\xi<z<R_g$, the third region is comprised of $R_g<z<2R_g$ and the last one in the center is representative for the bulk behavior. In Fig.~\ref{fig:gqz64_4M} we show the collective structure function, $G(q_\parallel)$, in these four intervals after the second stage of the packing process. We observe that $G(q_\parallel)$ in the interval from $0 < z < \xi$ deviates from its equilibrium distribution. This deviation is expected because our algorithm cannot reproduce local packing on the segmental scale.
The closer the slab is to the solid substrate, the larger are fluctuations at small $q_\parallel$. For distances $z>2R_g$, the fluctuations resemble the structure function of Fig.~\ref{fig:gq64_all} that has been obtained by averaging over the entire sample thickness.

\begin{figure}[tb]
\includegraphics[clip,width=1.\linewidth]{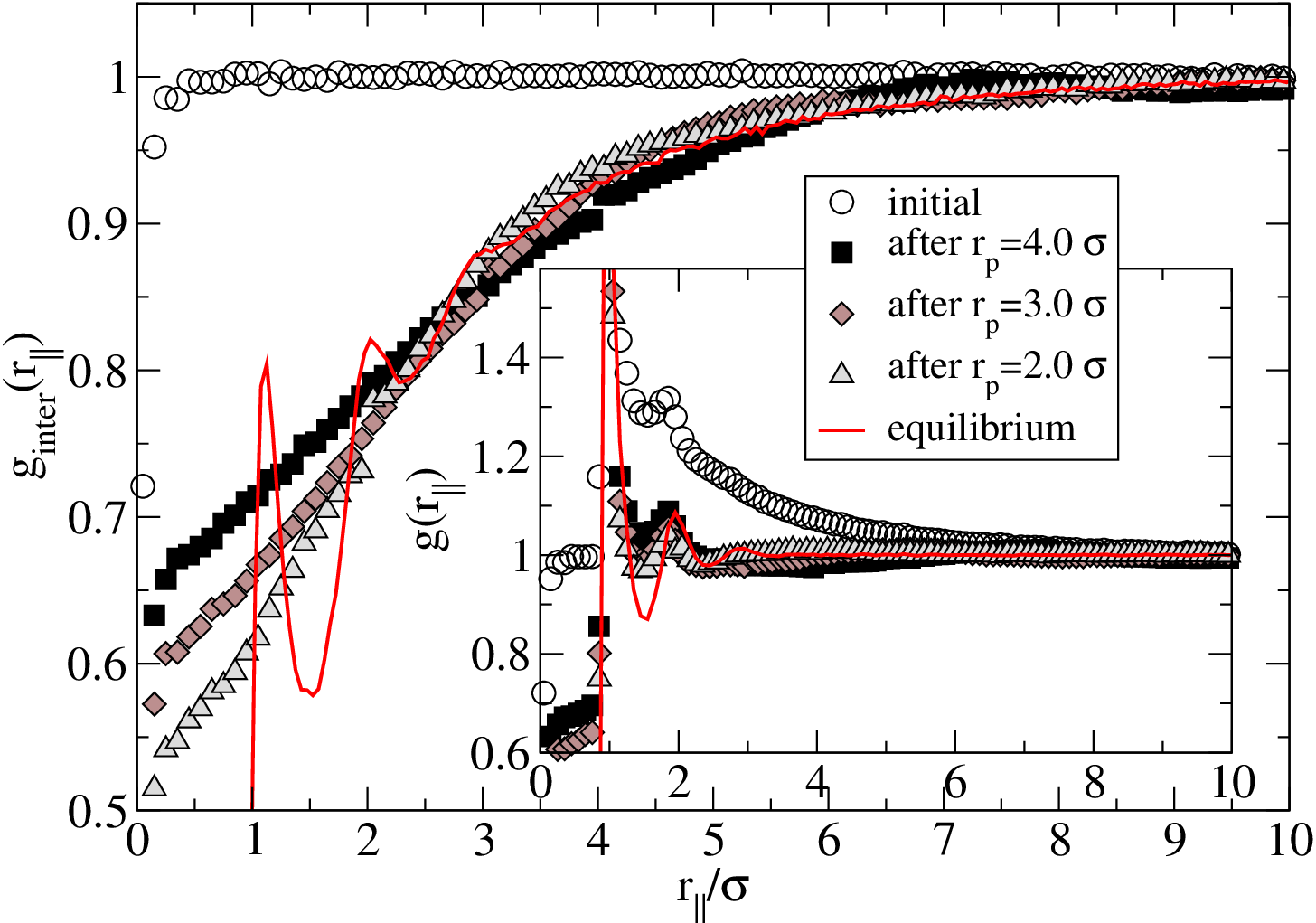}
\caption{\label{fig:pc64}The intermolecular pair correlation function $g_{\text{inter}}(r_\parallel)$ after each packing step for the system $N=64$ in comparison with the equilibrium distribution. The depth of the intermolecular part of the correlation hole is increasing with every packing step and on lengths larger than the $r_\parallel > 4\,\sigma$ the final distribution matches the equilibrium distribution very well. In the inset we present the total pair correlation function $g(r_\parallel)$ including the intramolecular contributions.}
\end{figure}

While the structure function measures the density fluctuations on different length scales it does not distinguish between intermolecular packing and intramolecular conformations. To quantify the distribution of distances between different chains, we calculated the intermolecular radial pair correlation function for segments within the same distance $\Delta z$ from the solid substrate
\begin{equation}
\rho g_{\text{inter}}(r_\parallel) = \frac{1}{nN} \sum\limits_{i\neq j}^n\sum\limits_{s,t=1}^N \Mv{\delta\left(\Abs{\mathbf{r}_{is}-\mathbf{r}_{jt}}-r_\parallel\right)\delta\left(z_{is}-z_{jt}\right)}
\end{equation}
as well as the total radial pair correlation function of the segments within the same distance from the solid substrate
\begin{equation}
\rho g(r_\parallel) = \frac{1}{nN} \sum\limits_{i,j=1}^n\sum\limits_{s,t=1}^N \Mv{\delta\left(\Abs{\mathbf{r}_{is}-\mathbf{r}_{jt}}-r_\parallel\right)\delta\left(z_{is}-z_{jt}\right)}.
\end{equation}
The former exhibits the intermolecular correlation hole while the latter is the Fourier transform of the structure function, $G$.
In the main panel of Fig.~\ref{fig:pc64}, we present $g_{\text{inter}}(r_\parallel)$ averaged over the entire system. After the first stage of the packing process (initial configuration), there are no intermolecular correlations, i.e., $g_{\text{inter}}(r) \equiv 1$. Already after the second stage of the packing with $r_p=4.0\,\sigma$, a pronounced correlation hole is established, and for $r_\parallel \gtrsim 5\,\sigma$, $g_{\text{inter}}(r_\parallel)$  matches the equilibrium distribution already very well. Upon decreasing the value of $r_p$ the correlation hole is getting deeper and $g_{\text{inter}}(r_\parallel)$ is approaching its equilibrium distribution. Two different regions can be distinguished: (i) For $r_\parallel\gtrsim 5\,\sigma$, the intermolecular correlations are determined by the packing of whole molecules and this effect is correctly reproduced by the packing algorithm. Hence, deviations from the equilibrium distribution are small. This is an important regime because creating this correlation hole on the scale of $R_g$ by a diffusive motion of the molecules would require an equilibration time of the order $\tau_R$. 
(ii) For $r_\parallel \ll 5\,\sigma$, the intermolecular correlations exhibit local packing which gives rise to an oscillatory behavior. For $r_\parallel\lesssim 1\,\sigma$, these local packing effects dominate the behavior and the $g_{\text{inter}}(r_\parallel)$ after the second stage deviates significantly from the equilibrium distribution. The length scale of the local packing effects is set by the size of a segment and the corresponding relaxation times are fast. For the rather small number of segments, $N=64$, the two regimes are not clearly separated and there is a rather gradual cross-over between the limiting behaviors that characterize the local structure of the fluid of segments and the polymeric correlation hole. 

The inset of Fig.~\ref{fig:pc64} depicts the total pair correlation function, $g(r_\parallel)$. At the beginning of the second stage of the packing process, $g(r_\parallel)$ exceeds unity for distances of the order $R_g$, i.e., $g(r_\parallel)-1$ is proportional to the intramolecular correlations. After the packing, however, $g(r_\parallel)$ has decayed to unity beyond $r_\parallel=5\,\sigma \approx 2\xi$ and the whole distribution closely resembles the equilibrium distribution.

\begin{figure}[tb]
\includegraphics[clip,width=1.\linewidth]{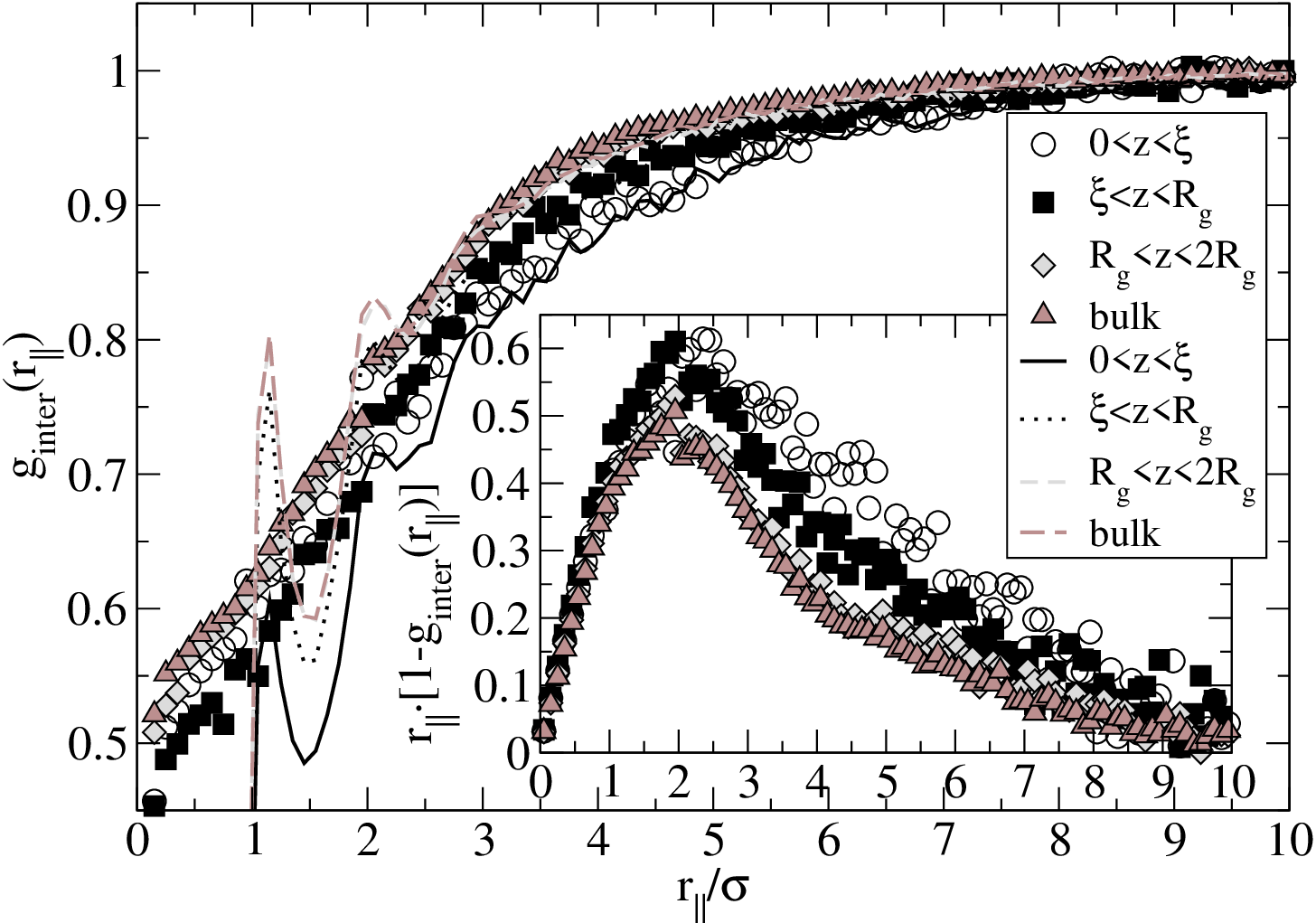}
\caption{\label{fig:pcz64_4M}$g_{\text{inter}}(r_\parallel)$ for $N=64$ after packing with $r_p=2.0\,\sigma$ for intervals with a different distance to the solid substrate. $g_{\text{inter}}(r_\parallel)$ shows in the interval $0<z<\xi$ a deeper correlation hole, that is reproduced by the packing algorithm. The inset shows $r_\parallel\cdot\left[1-g_{\text{inter}}(r_\parallel)\right]$ so that the correlation hole becomes clearly visible.}
\end{figure}

In Fig.~\ref{fig:pcz64_4M} we analyze the spatial correlations,  $g_{\text{inter}}(r_\parallel)$, in slabs with different distances from the solid substrate. The slabs are defined in the same way as in the discussion of the collective structure function. Although $g_{\text{inter}}(r_\parallel)$ shows in all regions a qualitatively similar behavior, the correlation hole near the solid substrates is considerably deeper and longer-ranged than in the bulk because of the back-folding of the chain conformations. On large scales ($r_\parallel \gtrsim 4\,\sigma$), this effect is perfectly reproduced by our packing algorithm.  In the inset of Fig.~\ref{fig:pcz64_4M}, we present $r_\parallel \cdot\left[1-g_{\text{inter}}(r_\parallel)\right]$ to highlight the dependence of the correlation hole on the distance to the solid substrate. 

\subsection{Fast relaxation}
\label{fast-relaxation}
\begin{figure}[tb]
\includegraphics[clip,width=1.\linewidth]{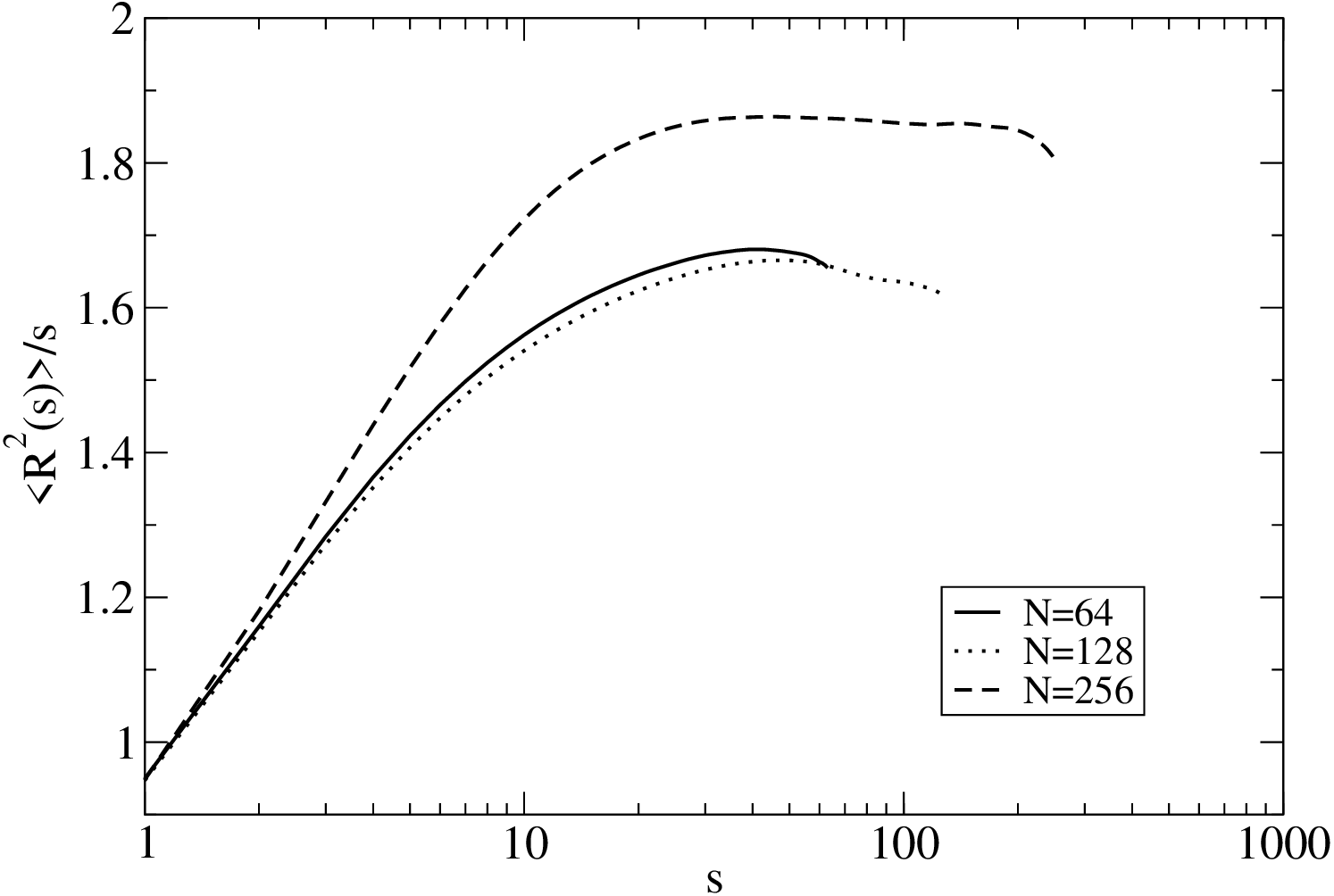}
\caption{\label{fig:R2s} $\mv{R^2(s)}/s\equiv \mv{R_\parallel^2(s)+R_\perp^2(s)}/s$, mean-squared size of chain segments of curvilinear length $s$ normalized to $s$ in equilibrium. }
\end{figure}

After having generated multi-chain configuration with the first two stages of our packing algorithm, we performed MD simulations to investigate the final equilibration during which the local packing structure of the fluid of segments is established. Due to the remaining overlap of segments, we use a push-off procedure (cf.~Sec.~\ref{model}) with the linearized Lennard-Jones potential of Eq.~\eqref{ljfc-potential} for non-bonded interactions. During these simulations, we monitored the collective structure function, $G(q_\parallel)$, the intermolecular pair correlation function, $g_{\rm inter}(r_\parallel)$, and the mean-squared size $\mv{R^2(s)}$ of chain segments of curvilinear length $s=\abs{n-m}<N$. Note that $R^2$ includes the parallel and perpendicular extension of the molecules in confinement. The latter quantity is averaged over all segments of size $s$, where $n$ and $m$ denote segment indices along the chain. Since $\mv{R^2(s)}$ is given by the integrated bond-bond-correlation function, it serves as a sensitive measure for intramolecular correlations.\cite{Wittmer04} For a Gaussian chain $\mv{R^2(s)}/s=b^2$, where $b$ is the statistical segment length. For chains in a melt, there are deviations from this law, particularly at small $s$. In Fig.~\ref{fig:R2s}, we have plotted $\mv{R^2(s)}/s$ in equilibrium for the three chain lengths under consideration. For $N=64$ and $N=128$, both distributions show similar behavior and they reach a plateau at $b^2$ for $s\to N$. The distribution for $N=256$ shows the plateau at a higher value of $b^2$, i.e. the chains are slightly more swollen because of the confinement.

\begin{figure}[tb]
\includegraphics[clip,width=1.\linewidth]{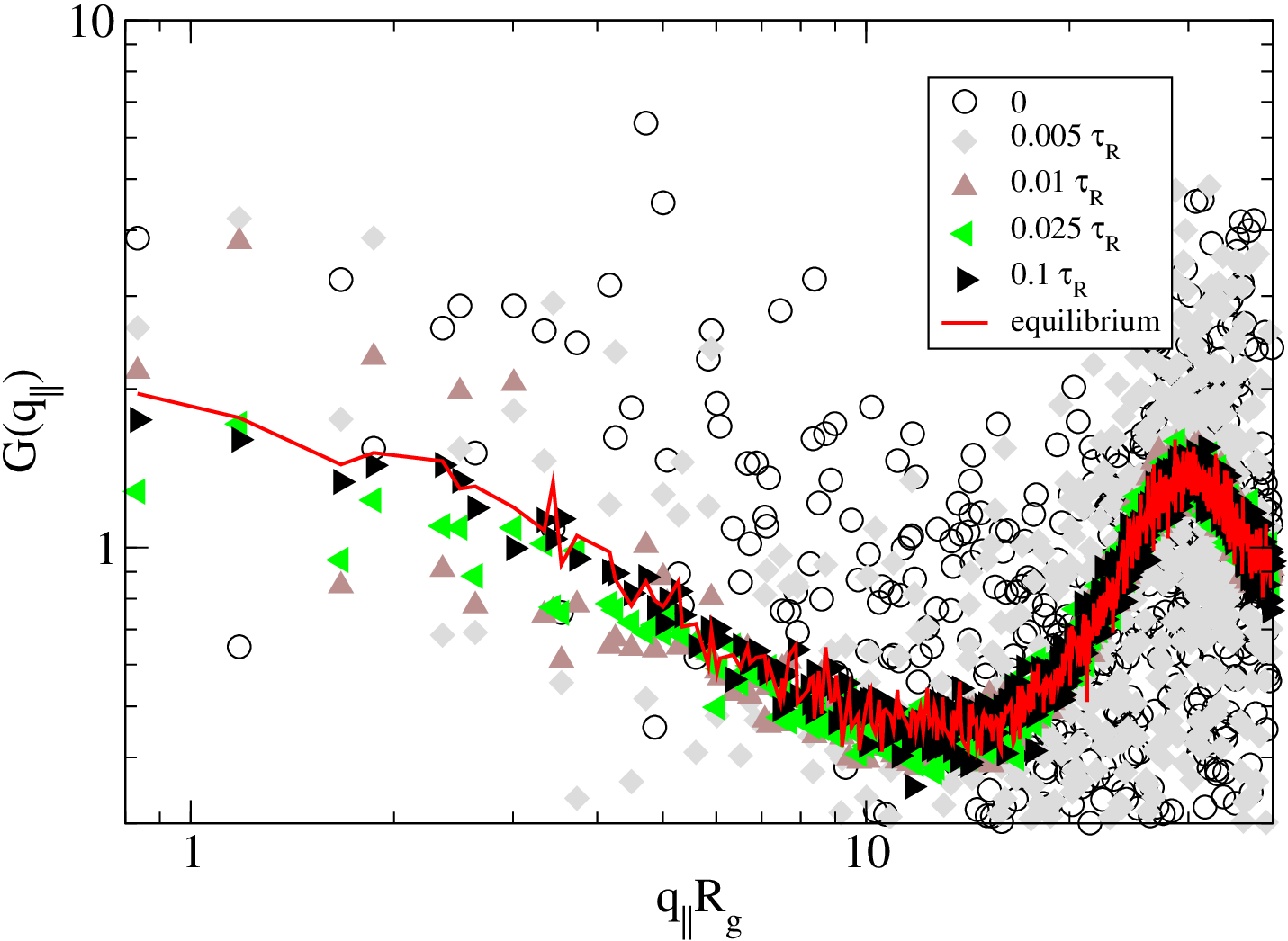}
\caption{\label{fig:gqzt64} $G(q_\parallel)$ for the slab $0<z<\xi$ at the beginning of the MD runs ($N=64$).}
\end{figure}

In our MD runs we observed that the relaxation on length scales of the size of a few bond lengths occurs very rapidly. Inspecting $G(q_\parallel)$, we notice that already after $\Delta t=10\,\tau$ there is hardly any difference compared to the equilibrium distribution visible. Even density fluctuations around $q_\parallel R_g \approx 7$ equilibrated completely, and also $g(r_\parallel)$ shows the same fast relaxation to equilibrium. In Fig.~\ref{fig:gqzt64} we present the fast equilibration of $G(q_\parallel)$ for the slab $0<z<\xi$, which has reached approximately the equilibrium distribution after $0.1\,\tau_R$.

\begin{figure}[tb]
\includegraphics[clip,width=1.\linewidth]{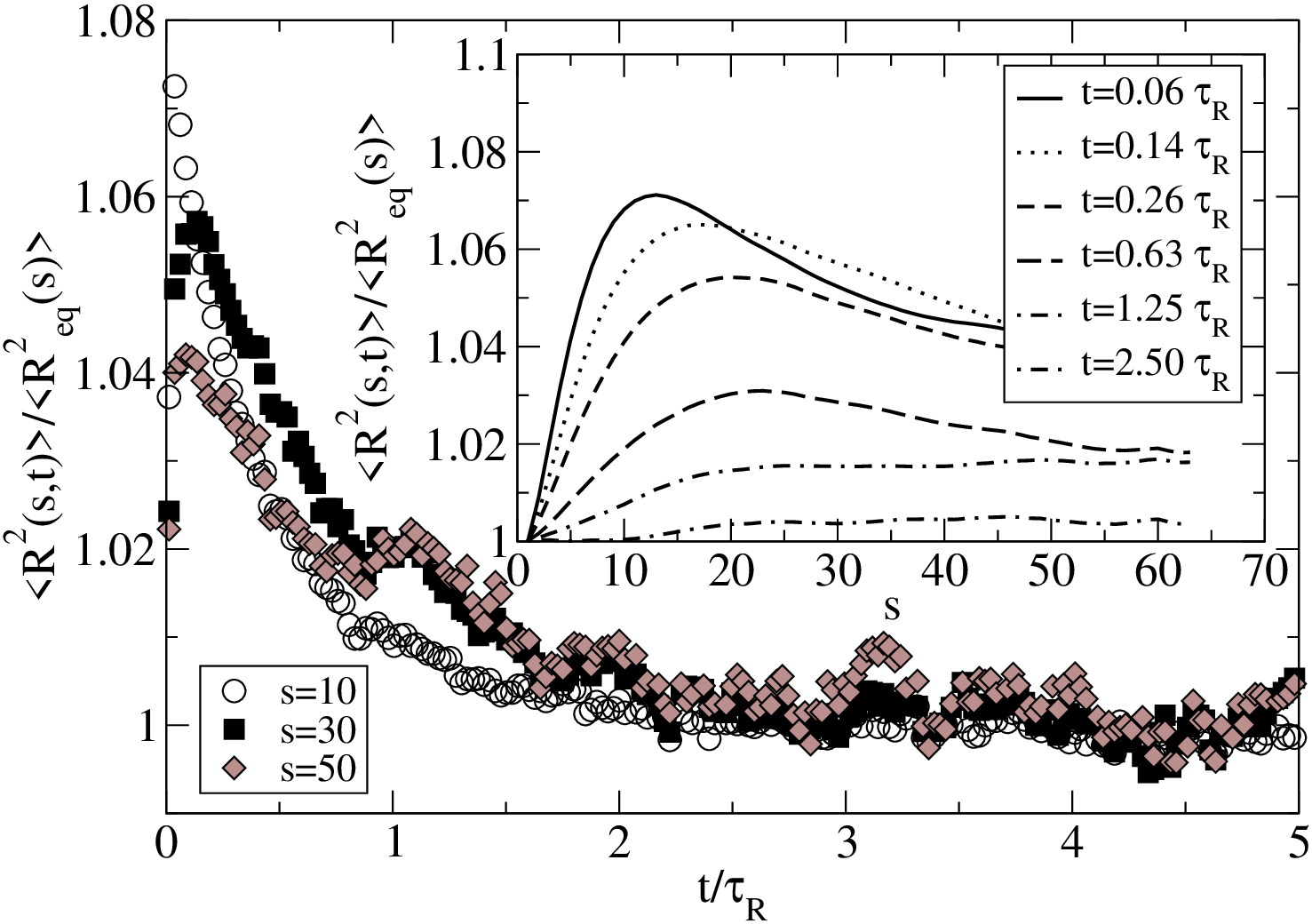}
\caption{\label{fig:idw64}The main panel presents $\mv{R^2(s,t)}/\mv{R^2_{\text{eq}}(s)}$ for $N=64$ as a function of time for a few fixed values of $s$, while in the inset the same quantity is depicted as a function of $s$ for various time steps.}
\end{figure}

Regarding $\mv{R^2(s)}$, however, it turns out that the intramolecular correlations have been disturbed and that their relaxation takes considerably longer than the relaxation of the local packing. The main panel of Fig.~\ref{fig:idw64} shows $\mv{R^2(s,t)}$ as a function of $t$ normalized by the equilibrium distribution $\mv{R^2_{\text{eq}}(s)}$ for $N=64$. The figure presents data for several values of $s$. $\tau_R$ is the Rouse time of a chain \cite{DoiEdwards} (cf.~Tab.~\ref{tab1}). The strongest deviations from equilibrium occur at $s\approx 15$ and they are smaller than $8\,\%$. Larger scales (larger values of $s$) show smaller deviations which decay exponentially in time. After roughly $1.5\,\tau_R$, the deviations are smaller than $1\,\%$ on all length scales and we consider the intramolecular correlations as being equilibrated. The inset of Fig.~\ref{fig:idw64} depicts $\mv{R^2(s,t)}/\mv{R^2_{\text{eq}}(s)}$ as a function of $s$ for fixed values of $t$. The peak of the strongest deviations moves slowly to larger $s$, importantly, it does however not propagate to the largest length scales.

\begin{figure}[tb]
\includegraphics[clip,width=1.\linewidth]{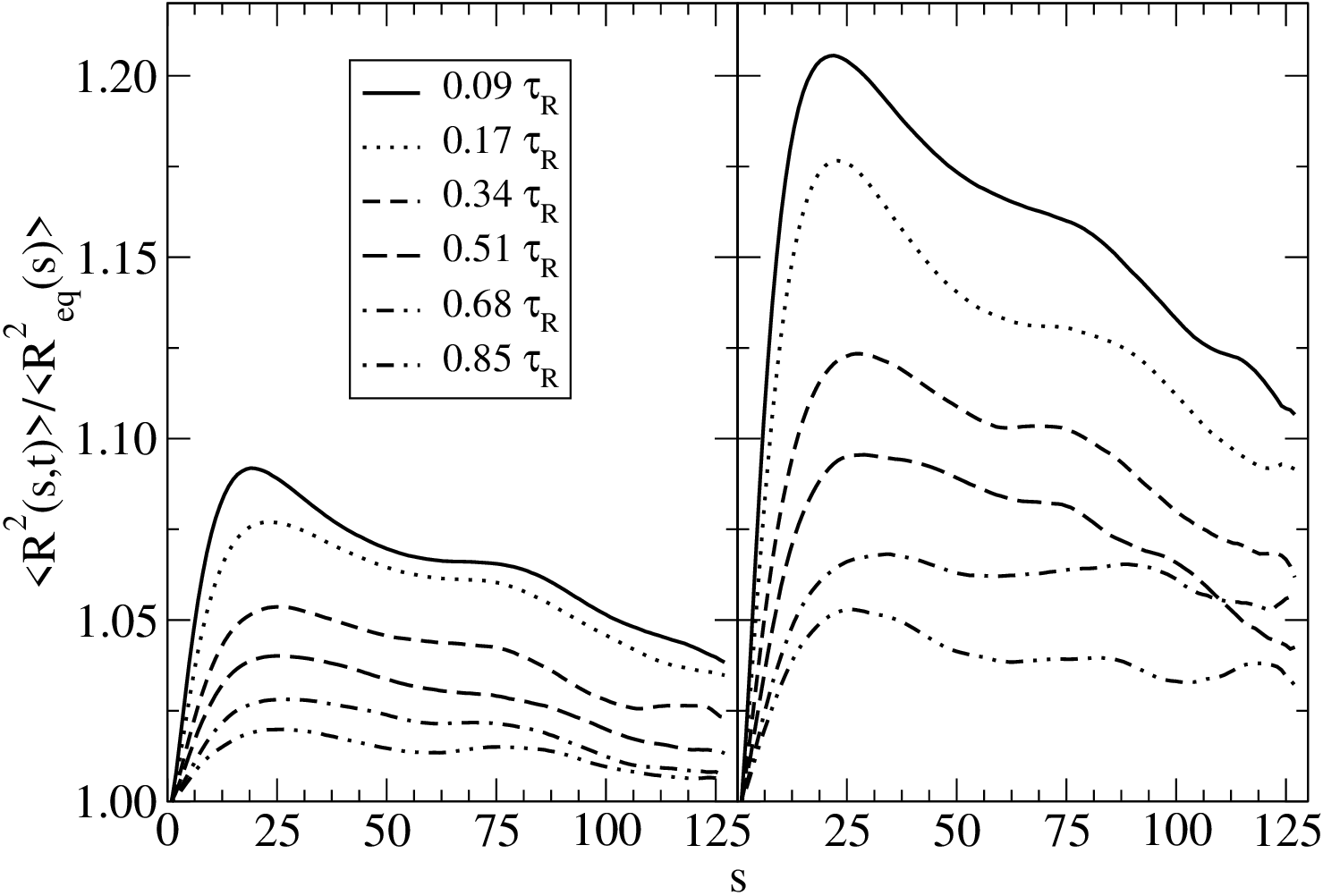}
\caption{\label{fig:vergleich}Comparison of the temporal evolution of $\mv{R^2(s,t)}/\mv{R^2_{\text{eq}}(s)}$ between an initial configuration generated by the packing algorithm (left) and a randomly packed configuration (first stage, right) for some fixed time steps $t$ ($N=128$).}
\end{figure}

In Fig.~\ref{fig:vergleich}, we finally compare the temporal evolution of $\mv{R^2(s,t)}/\mv{R^2_{\text{eq}}(s)}$ after the first stage of the packing with a configuration generated after the second stage. With a maximal deviation from the equilibrium distribution of $26\,\%$ at $s\approx13$ after the first stage, the intramolecular correlations are much stronger perturbed and also occur on slightly larger values of $s$ than in the configuration after the second stage. But even worse, these deviations in the configuration after the first stage propagate to larger length scales and therefore increase the required relaxation time significantly. 

\section{Conclusions and Outlook}
\label{discussion}
We have investigated the properties of a packing algorithm based on an idea by Auhl and co-workers \cite{Auhl03} that arranges given single-chain conformations extracted from an inhomogeneous polymer melt in order to generate a multi-chain configuration, which can serve as a starting configuration for a molecular dynamics simulation. The data demonstrate that reducing local density fluctuations, which are quantified by the average number of segments that surround a reference bead, we are able to re-establish intermolecular correlations on the length scale of the molecules' extension, $R_e$. Most prominently, the packing algorithm generates the correlation hole in the intermolecular pair correlation function.

In order to assess the quality of the packing algorithm, we utilize single-chain conformations extracted from an equilibrated thin polymer film. The so-generated single-chain properties are exact by construction. In practice, mean-field techniques that evaluate the single-chain partition function via a partial enumeration over a large ensemble of equilibrated single-chain conformations of the bulk will provide an excellent approximation for the single-chain properties in an inhomogeneous system. Since the single-chain conformations are extracted from the bulk they already incorporate the long-ranged intramolecular correlations that the incompressibility constraint imparts on the conformations (e.g., the power-law decay of the bond-bond correlation function along the molecule).

The local fluid-like packing structure on the length scale of the excluded volume screening length is not reproduced by the packing algorithm. This local structure, however, can be established on a short time scale compared to the relaxation time of the inter- and intramolecular correlations on the scale $R_e$. Thus, the algorithm is particularly useful for creating initial configurations for particle-based simulations of inhomogeneous polymer melts or semi-dilute solutions on the basis of mean-field results.

There remain two problems to be addressed: 
(i) For inhomogeneous melts of very long chains, even the generation of equilibrated bulk configurations is a computational challenge.\cite{Auhl03} Gaussian chain conformations, however, are inappropriate because the long-ranged intramolecular correlations require protracted long times to equilibrate. Instead of drawing long polymers directly from an equilibrated melt it might be feasible to generate appropriate conformations by joining short chains \cite{A69,Kremer88} accounting for the weak effective intramolecular interactions \cite{DoiEdwards} due to the incompressibility.  This proposed procedure is very similar to the analytical perturbation calculations performed by \citet{Semenov03} and we  expect it to reproduce the non-trivial long-ranged intramolecular correlations. \cite{Wittmer04} These single-chain conformations can be used as a starting point for mean-field calculations of inhomogeneous systems and the results of the latter serve as input into the packing algorithm. 

(ii) We have applied the packing algorithm to a polymer melt with the simplest inhomogeneity, i.e. a thin polymer film confined by symmetric solid substrates. We emphasize that the packing procedure can be readily generalized to more complicated geometries. The moves, which are utilized to arrange the single-chain conformations during the packing process, must not alter the single-chain properties (i.e., the intramolecular structure and the position and orientation with respect to the spatial inhomogeneity). Thus, in the case, where translational invariance is broken in all three spatial directions, only replacement moves are acceptable but no translations or rotations.

\begin{acknowledgments}
We thank D. Bedrov, G. D. Smith, and J. P. Wittmer for useful discussions. Computing time was generously provided by the GWDG G{\"o}ttingen, RRZN Hannover and HLRN Hannover/Berlin, Germany. Financial support by the Volkswagen foundation and the DFG under grant Mu 1674/9 are gratefully acknowledged.
\end{acknowledgments}


\end{document}